\documentclass[
superscriptaddress,
amsmath,amssymb,
aps,
prl,
twocolumn
]{revtex4-1}

\usepackage{wrapfig}

\usepackage{tikz}
\usepackage{graphicx}
\graphicspath{{figures/}}
\usepackage{dcolumn}
\usepackage{bm}
\usepackage{hyperref}
\usepackage{multirow}
\usepackage{array}
\usepackage{booktabs}
\usepackage{ctable}
\usepackage{upgreek}
\usepackage{epsfig,psfrag,subfigure,amsopn}
\usepackage{mathrsfs}
\usepackage{amssymb}
\usepackage{amsbsy}
\usepackage{color}
\usepackage{cancel}
\usepackage{}

\usepackage{tcolorbox}

\usepackage{pifont}
\usepackage{marginnote}
\usepackage{float}
\newcommand{\bcen}{\begin{center}}
	\newcommand{\ecen}{\end{center}}
\newcommand{\btab}{\begin{tabular}}
	\newcommand{\etab}{\end{tabular}}
\newcommand{\bdes}{\begin{description}}
	\newcommand{\edes}{\end{description}}

\newcommand{\beq}{\begin{equation}}
\newcommand{\eeq}{\end{equation}}
\newcommand{\bea}{\begin{eqnarray}}
\newcommand{\eea}{\end{eqnarray}}

\newcommand{\bary}{\begin{array}}
	\newcommand{\eary}{\end{array}}
\newcommand{\benum}{\begin{enumerate}}
	\newcommand{\eenum}{\end{enumerate}}
\newcommand{\bitem}{\begin{itemize}}
	\newcommand{\eitem}{\end{itemize}}

\makeatletter

\newcommand{\Rmnum}[1]{\expandafter\@slowromancap\romannumeral #1@}
\makeatother

\newcommand{\sectionprl}[1]{{\em #1}\/.---}

\newcommand{\titlename}{Gap Statistics for Confined Particles with Power-Law Interactions}

\begin{document}
	
	\title{\titlename}

\author{S. Santra}
\email{saikat.santra@icts.res.in} 
\address{International Centre for Theoretical Sciences, Tata Institute of Fundamental Research, Bengaluru -- 560089, India}

\author{J. Kethepalli}
\email{jitendra.kethepalli@icts.res.in}
\address{International Centre for Theoretical Sciences, Tata Institute of Fundamental Research, Bengaluru -- 560089, India}

\author{S. Agarwal}
\email{sanaa.agarwal@colorado.edu} 
\address{Department of Physics, University of Colorado, Boulder, Colorado 80309, USA}
\author{ A. Dhar}
\email{abhishek.dhar@icts.res.in}
\address{International Centre for Theoretical Sciences, Tata Institute of Fundamental Research, Bengaluru -- 560089, India}

\author{M. Kulkarni}
\email{manas.kulkarni@icts.res.in}
\address{International Centre for Theoretical Sciences, Tata Institute of Fundamental Research, Bengaluru -- 560089, India}

\author{A. Kundu}
\email{anupam.kundu@icts.res.in}	
\address{International Centre for Theoretical Sciences, Tata Institute of Fundamental Research, Bengaluru -- 560089, India}


	\date{\today}
	
\begin{abstract}
We consider the $N$ particle classical Riesz gas confined in a one-dimensional external harmonic potential  with power-law interaction of the form $1/r^k$, where $r$ is the separation between particles. As special limits it contains several systems such as  Dyson's log-gas ($k\to 0^+$), Calogero-Moser model ($k=2$), 1D one component plasma ($k=-1$) and the hard-rod gas ($k\to \infty$). Despite its growing importance, only large-$N$ field theory and average density profile are known for general $k$. In this Letter, we study the fluctuations in the system by looking at the statistics of the gap between successive particles. This quantity is analogous to the well-known level-spacing statistics which is ubiquitous in several branches of physics. We show that 
the variance goes as $N^{-b_k}$ and we find the $k$ dependence of $b_k$ via direct Monte Carlo simulations. We provide supporting arguments based on microscopic Hessian calculation and a quadratic field theory approach.  We compute the gap distribution and  study its system size scaling. Except in the range $-1<k<0$, we find scaling for all $k>-2$ with both Gaussian and non-Gaussian scaling forms.  \end{abstract}	
	
\maketitle

\sectionprl{Introduction}
The Riesz gas, consisting of $N$ particles with long-range interactions confined in a harmonic trap, is one of the classic examples of a strongly interacting many-body system. The model is characterized by power-law interactions potentials of the form $V(r) \sim J r^{-k}$, where $r$ is the distance between two particles, $J>0$ is the interaction strength, and $k>-2$ (to ensure stability). Special values of $k$ lead us to some important models such as the log-gas ($k \to 0^+$)~\cite{Dyson_JMP_1962,dyson_JKP_1962}, the one-dimensional one-component plasma (1dOCP, $k=-1$)~\cite{Lenard_JMP_1963,Statistical_MPCPS_Baxter,dhar_IOP_2018,dhar_PRL_2017} and the Calogero-Moser (CM) model ($k=2$)~\cite{Calogero_JMP_1969,Calogero_JMP_1971,Calogero_LNC_1975, polychronakos_IOP_2006}. Experimental realizations of this model in cold atom systems have now become possible~\cite{brown_CUP_2003,chalony_PRA_2013,zhang_nature_2017} and hence it is essential to have a complete characterization of  its equilibrium and dynamical properties. The long-range nature of the interactions makes this difficult but some progress has recently been cmade~\cite{sanaa2019,Dean_EPL_2019,dhar_IOP_2018,dhar_PRL_2017,Bun_PRL_2014,Majumdar_IOP_2014,Lu_PRL_2011,Dean_PRE_2008,Dean_PRL_2006,GUSTAVSSON_2005,Cunden_JSM_2017,Rojas_PRE_2018,Cunden_Springer_2019}. In Ref.~\cite{sanaa2019} the exact density profile was computed using a field theoretic approach, thereby obtaining a generalization of the Wigner semicircle law for the log-gas~\cite{Wigner_MPCPS_1951}.  The form of the average density profile and the scaling of its support with increasing $N$ was found to be nontrivial. For the 1D one-component plasma for which the density profile is flat, the  distribution of the position of the right most particle was computed exactly~\cite{dhar_PRL_2017} and found to be different from the Tracy-Widom form~\cite{Tracy_CMP_1994,Tracy_CMP_1996}. Surprisingly, the density profile in the CM gas is identical to that in the log-gas but the edge particle distribution takes a different (non  Tracy-Widom) form~\cite{sanaa2019JSP}. Recently, the average density profile, in the presence of a hard wall, has also been computed exactly for all $k>-2$~\cite{kethepalli_JSM_2021}.

One of the interesting observations of Ref.~\onlinecite{sanaa2019} was  on the system-size scaling of the mean separation $ \langle \Delta \rangle$ between neighboring particles. This has the form  $ \langle \Delta \rangle \sim N^{-a_k}$ where $a_k$ has a nonmonotonic dependence on $k$ and can have both positive and negative signs. 
For a complete characterization it is necessary to go beyond the mean and study the fluctuations of this quantity as well as its full distribution.  The interplay between the long-range interactions and the confining potential makes this a fascinating and difficult question and this is the main focus of this Letter. 

The gap statistics is analogous to level spacing statistics which has been studied in great detail in different areas such as random matrix theory (RMT)~\cite{mehta2004random,forrester_PUP_2010} and quantum chaos~\cite{Berry_PRS_1977,Bohigas_PRL_1984,IZRAILEV_physics_1990,haake_quantum_1991}. In the context of RMT we recall that the equilibrium distribution of particle positions in the log-gas ($k\to 0^+$, $J\to \infty$, with $Jk \to J_0$ ) at inverse temperature $\beta$ corresponds to the distribution of eigenvalues of random matrices for the Gaussian orthogonal (GOE), unitary (GUE), and symplectic (GSE) ensembles, corresponding to Dyson indices $1,2,$ and $4$ respectively. From  this correspondence it is known that the distribution of particle spacing, normalized by the mean spacing, is given quite accurately by the Wigner surmise (WS)~\cite{mehta2004random,Wigner_MPCPS_1951,forrester_PUP_2010}.
A variant of the WS has also been applied to the CM model ($k=2$)~\cite{bogomolny_PRL_2009} but to the best of our knowledge, there are no results for other values of $k$ and this Letter provides a complete characterization. Needless to mention, fluctuations at the microscopic level is an avenue that is essentially unexplored in systems with long-range interactions. Probing such fluctuations has now become experimentally accessible given the recent breakthroughs in the technology of quantum gas microscopy~\cite{Bakr_Springer_2009,Lawrence_PRL_2015,Haller_nature_2015,Parsons_PRL_2015,Kush_NSR_2016,gross_nature_2021}. Gap fluctuations give us a novel way to probe aspects of the underlying interacting systems that are otherwise completely elusive to diagnostics such as density profiles.

\begin{figure}[t]
	\begin{minipage}[H]{0.4\textwidth}
		\centering		\includegraphics[width=\textwidth]{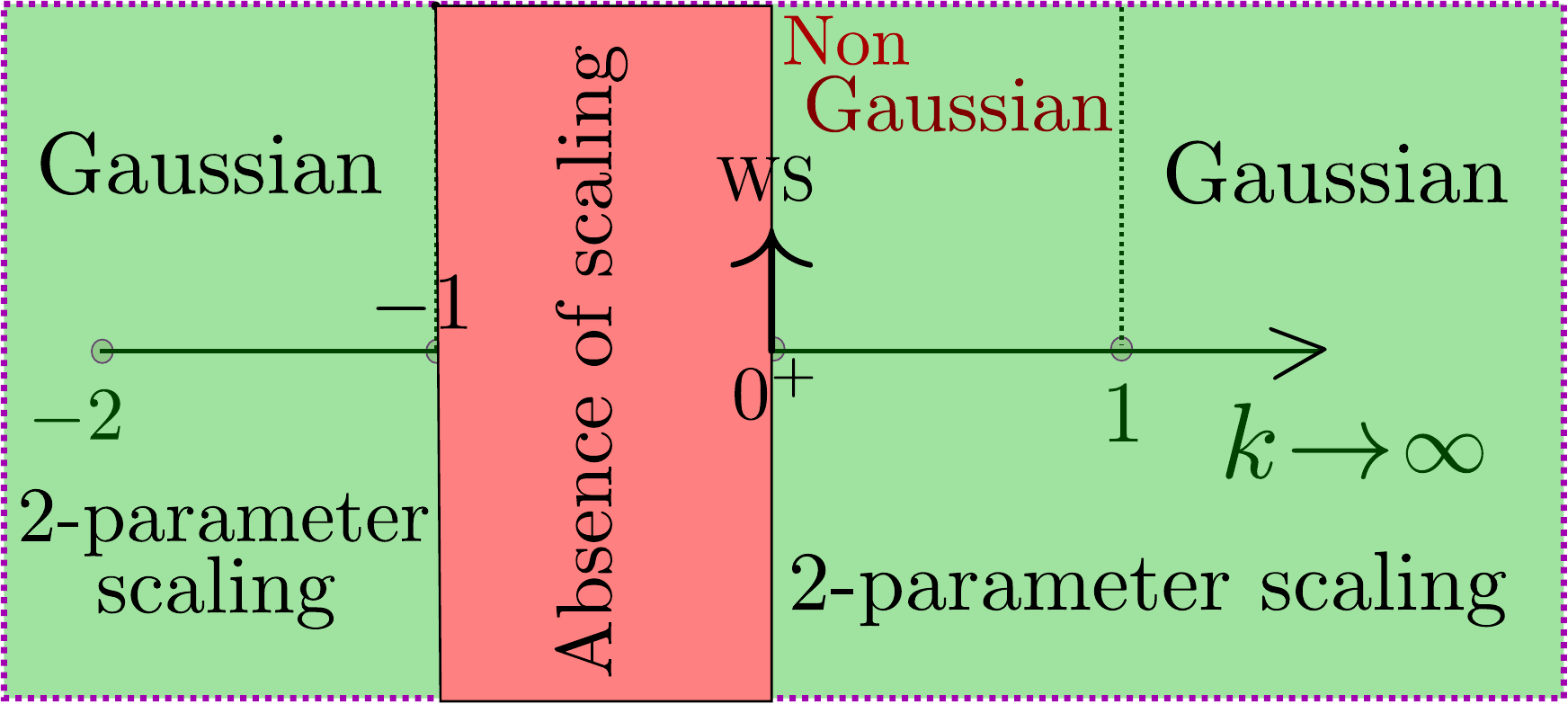}		
	\end{minipage}

	\caption{Schematic phase diagram of the behaviour of the gap distribution. We find four regimes in $k \in (-2,\infty)$, where the gap distribution has different scaling properties. In the region $(-2,-1) \cup (0,\infty)$ the scaling limit is achieved by using 
	mean and variance of gap only. The scaling function for $k \in (-2,-1) \cup (1,\infty)$ is Gaussian whereas it  is non-Gaussian  in $k \in (0,1)$. In the regime $k \in (-1,0)$ we are unable to obtain a scaling limit.}
		\label{fig:summary}
\end{figure}

 Our main results are the following: 
(i) From direct Monte Carlo (MC) simulations, we find that the system size scaling of the variance of the bulk gap is characterized by a non-trivial exponent $b_k$ that fits the form in 
Eq.~(\ref{eq:bk}). (ii) This proposed form is further validated from our results based on a microscopic Hessian (MH) calculation and a quadratic field theory (FT). 
(iii) We study the scaling properties of the gap distributions for different $k$ and observe that there exists four regimes as shown in Fig.~\ref{fig:summary}.


\sectionprl{Model and definitions}
The harmonically confined Riesz gas consists of $N$ classical particles, confined in a harmonic potential on a line and interacting with each other via pairwise repulsion. We denote the positions of the particles on the line by ${x_i} \,(i=1,2,...,N)$. The pairwise repulsive interaction is taken as a power law of the distance between the particles, and the total potential energy is given by ($ \forall k > -2$) \cite{riesz_ASU_1938} \\
\bea
E(\{x_j\})=\sum_{i=1}^{N} \frac{x^2_i}{2}+ \frac{J ~\text{sgn}(k)}{2} \sum_{i\neq j} \frac{1}{\mid x_i-x_j \mid ^k}, 
\label{eqn:1}
\eea
where $J>0$ and  $\text{sgn}(k)$ ensures a repulsive interaction. We consider a thermal distribution of the $N$ particles given by
$P_{G}(x_1,x_2,\ldots,x_N)={e^{-\beta E}}/{Z}$, where $Z$ is the partition function and henceforth we set the inverse temperature $\beta=1$. Without loss of generality, we assume that the particles are ordered, i.e.,  $x_1\leq x_2 \leq x_3 \ldots \leq x_N$. The mean thermal density of particles is defined as
$ \rho^{\rm (eq)}_N(x) = (1/N) \sum_{i=1}^N \langle \delta(x-x_i) \rangle$ ,
where $\langle ...\rangle$ denotes a thermal average over the distribution $P_G(\{x_i\})$. 
 For large $N$ the average density has been computed exactly for all $k$ values~\cite{sanaa2019} and has a finite support in the range $[-l_k N^{\alpha_k}/2, l_k N^{\alpha_k}/2]$ (for $k \neq 1$ \footnote{Note that for the marginal case  $k=1$ the support scales as
$\sim (N \ln N)^{1/3}$.})     
where the exponent $\alpha_k={k}/{(k+2)}$  for $ k>1$ and ${1}/{(k+2)}$ for  $-2<k<1$, with $l_k$ known explicitly~\cite{sanaa2019,SM}. 
 The average density $\rho^{\rm (eq)}_N(x)$ for large $N$ and temperature $T < N^{2\alpha_k}$ is given by the scaling form $
 \rho^{\rm (eq)}_N(x) = {(l_k N^{\alpha_k})}^{-1} F_k \left({x}/{(l_k N^{\alpha_k})}\right)$, where the scaling function $F_k (y)$ is known exactly~\cite{sanaa2019}.

The main quantity of interest here is the interparticle separation $\Delta_i=x_{i+1}-x_i$ and the normalized separation $s_i= \Delta_i/ \langle \Delta_i \rangle$. The distribution of $s$ is defined as
\begin{align}
\label{sdist}
P_N(s)= \frac{1}{N-1} \sum_{i=1}^{N-1} p^{(i)}_N(s), 
\end{align}
where $p^{(i)}_N(s)= \langle  \delta (s - s_i) \rangle$ is the distribution of the $i$-th normalized gap. We expect that for typical fluctuations, $P_N(s)$ will be dominated by the bulk gaps, but edge gap contributions could be  important for atypical $s$.

\begin{figure}[t]
		\includegraphics[width=0.47\textwidth]{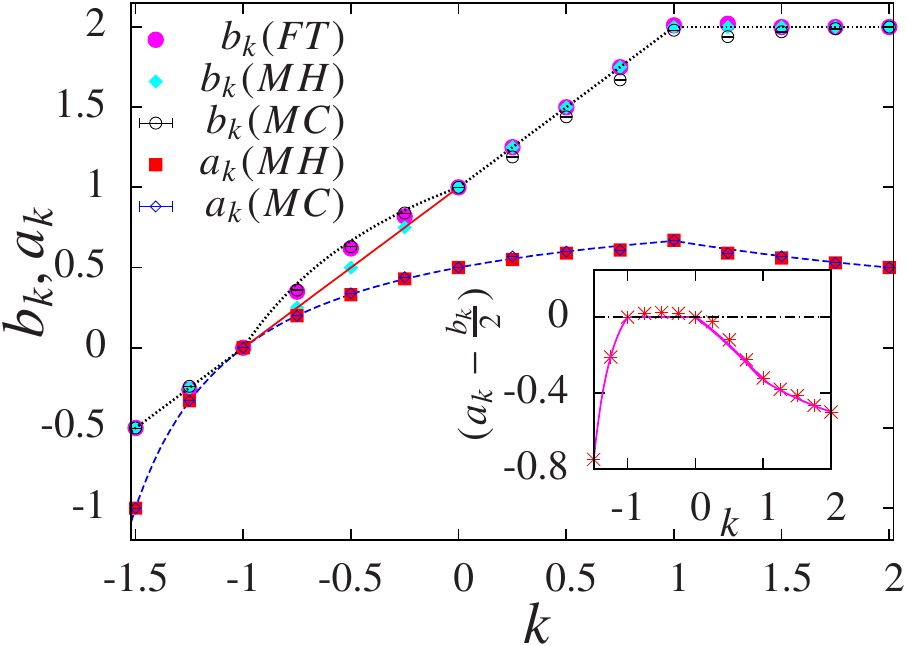}		

	\caption{Comparison of the exponents $a_k$ and $b_k$ (symbols) obtained from simulations (MC), from MH and FT calculations, with  Eqs.~\eqref{eq:ak} (dashed line) and \eqref{eq:bk} (dotted line).  In the inset we plot $\big (a_k-\frac{b_k}{2} \big )$ which quantifies the relative fluctuations $\sigma_{\Delta}/ \langle \Delta \rangle $ in the MC data. To extract $b_k$ the largest system sizes used were $N=2048$ for MC, $16384$ for both MH and FT. All the errorbars are smaller than the point symbols.}
		\label{fig:akbk_k}
\end{figure}

\sectionprl{Results for mean and variance of bulk gap} We expect that for bulk particles $1 \ll i \ll N-1$, the average bulk gap should scale as $\langle \Delta_{i} \rangle \sim N^{\alpha_k}/N = N^{-a_k}$, where $a_k=1-\alpha_k$, i.e., 
\bea
\label{eq:ak}
a_k=	 \begin{cases} \frac{2}{k+2} & \text{for } k>1 \\
	\frac{k+1}{k+2} &\text{for}   -2<k<1.
	 \end{cases}	
\eea
We also expect a power law dependence on the system size of the gap fluctuations $\sigma^2_{\Delta_i}=\langle \Delta_i^2 \rangle - \langle \Delta_i \rangle^2$.  In particular,  for the mid-gap corresponding to $i=N/2$,  we provide theoretical arguments based on MH and FT (see later) for the  conjecture:
 \begin{equation}
\sigma^2_{\Delta_{N/2}} \sim N^{-b_k},~~\text{where},
\end{equation} 
\begin{align} 
\begin{split}
b_k  &= \begin{cases} 2 ~~ &{\rm{for}}~~ k>1 \\
	1+k & {\rm for}  ~ 0<k<1 \\
	2(k+1)/(k+2) & {\rm for}  ~ -1<k<0  \\ 
	1+k  ~~   &{\rm for}  ~~ -2<k<-1. \end{cases}
\end{split}
\label{eq:bk}
\end{align}
\begin{figure}[t]
	\begin{minipage}[H]{0.48\textwidth}
		\centering		\includegraphics[width=\textwidth]{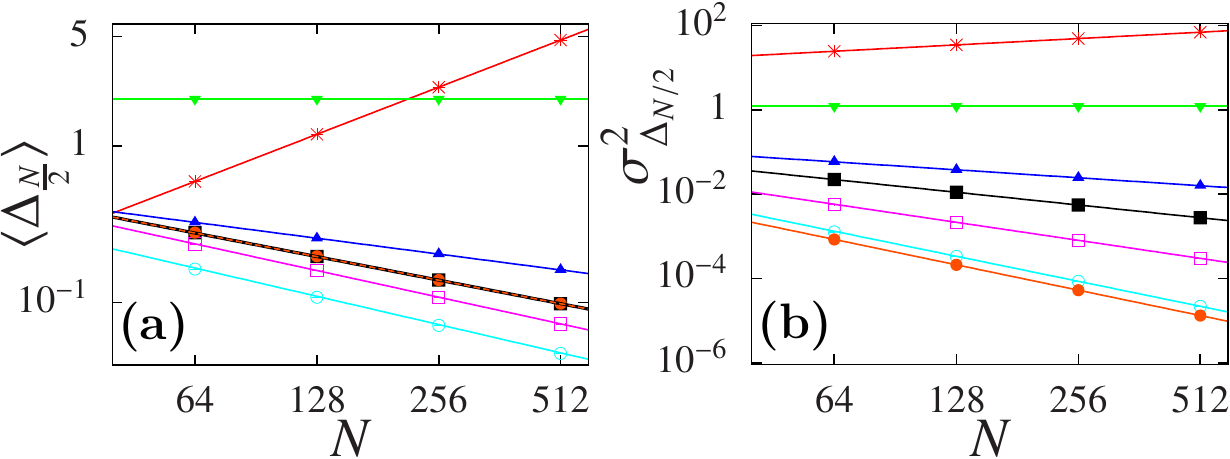}		
	\end{minipage}
	\caption{Mean (left) and variance (right) of mid gap as a function of system size for $k=2$(orange filled circle), $1.5$(blue circle),$0.5$(red square), $0^+$(black filled square), $-0.5$(blue upward triangle), $-1$(green downward triangle) and $-1.5$(red asterisk). Solid lines correspond to their corresponding power-law fitting  (Eq.~\ref{eq:bk}). The slopes in (a) are $a_k=0.5,0.57,0.6,0.5,0.33,0,-1$  and in (b) are $b_k=2,1.97,1.42,1,0.63,0,-1$,   for decreasing $k$. These are consistent with Eq.~\eqref{eq:ak} and Eq.~\eqref{eq:bk} as elucidated in Fig.~\ref{fig:akbk_k}. The error bars are negligible~\cite{SM}. In (a)  the data for $k=-1.5$ is scaled by a factor $500$. $\sim10^8$ MC samples are used for the computations.}
	\label{fig:mean_variance}
\end{figure}
We present numerical evidence for the above conjecture in Figs.~\ref{fig:mean_variance} and \ref{fig:akbk_k} where we observe reasonable agreement between the numerically obtained exponent (MC) and the conjectured values. We believe that the slight deviations from the predictions for few values of $k$ are due to finite size effects, since the error bars are small (see Ref.~\cite{SM} for discussion of error bars). 
We verified that the above scaling in Eq.~\eqref{eq:bk} also holds for other gaps deep in the bulk. Interestingly we find that for $-1 \leq k \leq 0$, the ratio $\sigma_{\Delta_i}/ \langle \Delta_i \rangle$ as well as $P^{(i)}_N(s)$ are weakly dependent on $i$ (for large $N$ and $i$ in the bulk; see Sec.~III of Ref.~\cite{SM}).

\begin{figure}[t]
	\begin{minipage}[H]{0.48\textwidth}
		\centering		\includegraphics[width=\textwidth]{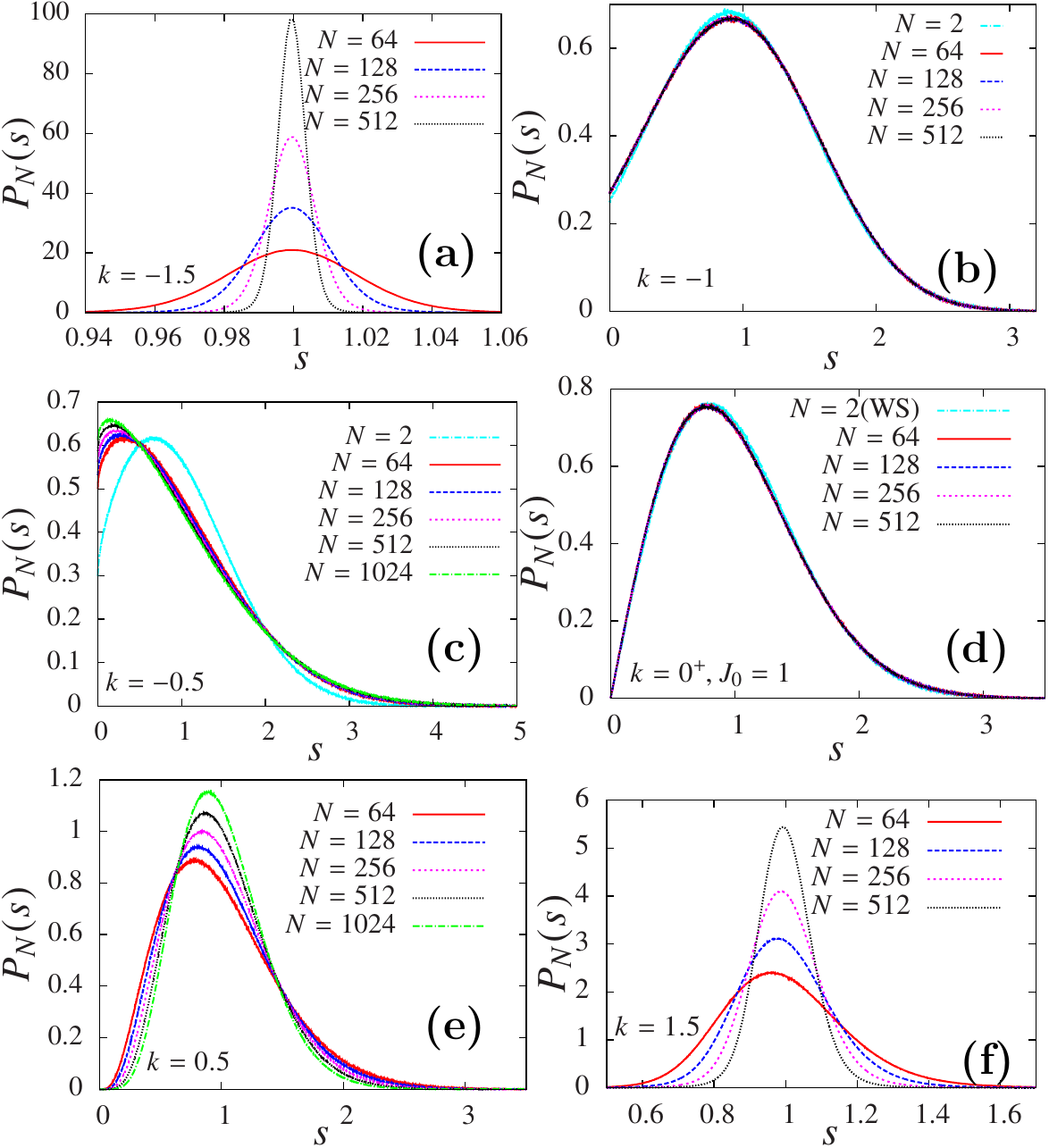}		
	\end{minipage}
	\caption{Plot of distributions $P_N(s)$ for different values of $k$ and $N$. Except for the  and 1DOCP (b) and log-gas (d),  for other values of $k$, we do not see convergence in $N$ which naturally implies that it is different from $P_2(s)$ and hence there is no generalisation of WS. }
	\label{no-ws-for-gen-k}
\end{figure}

\begin{figure}[t]
	\begin{minipage}[H]{0.49\textwidth}
		\centering		\includegraphics[width=\textwidth]{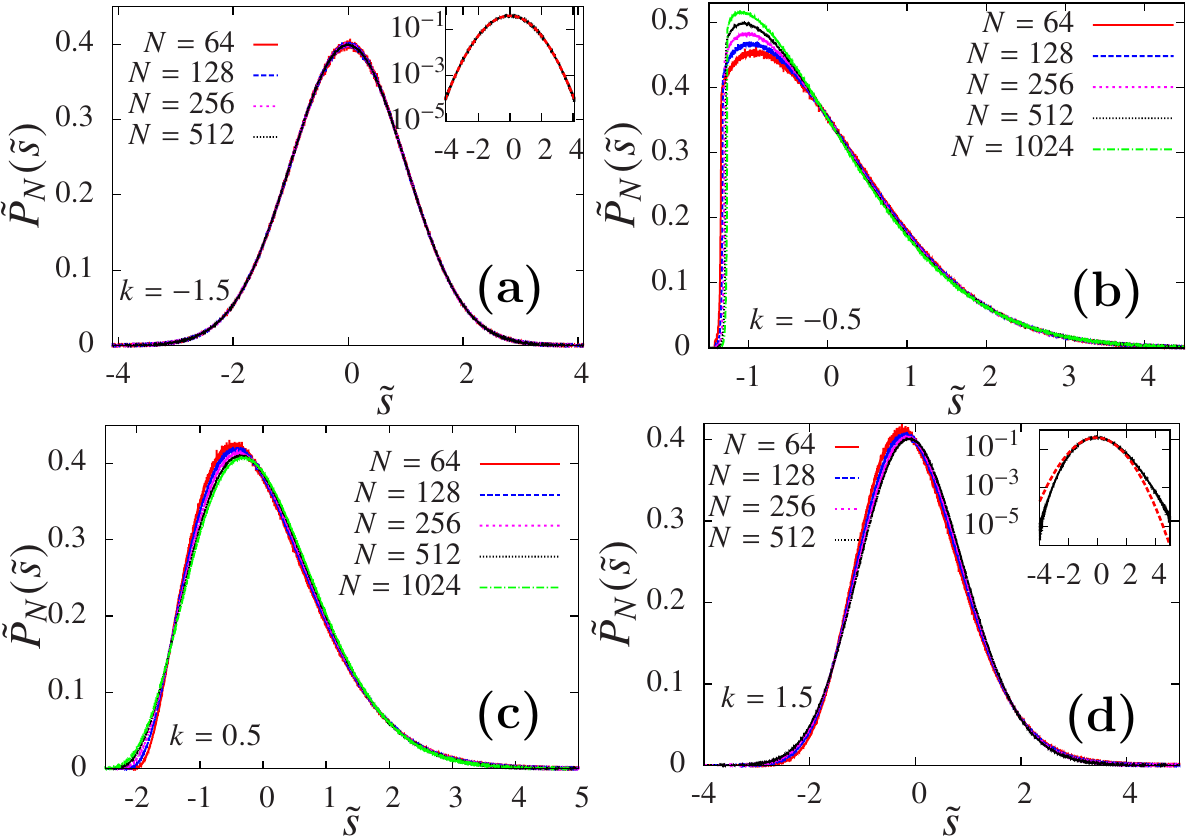}		
	\end{minipage}
	\caption{Plot of $\tilde{P}_N(\tilde{s})$ for different values of $k$ and $N$. The distributions for $k=\pm 1.5$ are fitted with a Gaussian over two standard deviations (insets). This is generally observed for $k \notin [-1,1]$. 
	The distributions for $k=\pm 0.5$ are very different from Gaussian and this is generally the case for $k \in [-1,1]$.}
	\label{tildeP_n(s)}
\end{figure}

\sectionprl{Results for distribution of gap}: The distribution of the normalized gap $s$ in Eq.~\eqref{sdist} is a well studied object in RMT~\cite{sorensen_Springer_1991,poli_PRL_2012,atas_APS_2013,allgaier_PRE_2014,wang_PRB_2020} where one of the important results is on the universal form of $P_N(s)$ given by the WS. For the distribution of eigenvalues of the random matrices belonging to the three Gaussian ensembles, with Dyson indices $1,2,4$ (which for our log-gas corresponds to $\beta J_0=1,2,4$), it is known that $ P_{N \to \infty} (s)$ is in fact accurately described by $P_{2}(s)\equiv P_{N=2}(s)$ (which is basically the WS) and is given by \cite{mehta2004random,Wigner_MPCPS_1951}
$P_2(s)= A_0 s^{\beta J_0} e^{-B_0 s^2},~~(\text{for~log~gas}),\label{ws-k->0}
$
where $A_0$ and $B_0$ are constants. From our simulations we in fact find that the WS for the log-gas is quite accurate for all $\beta J_0 >1$.  We now examine the distribution $P_N(s)$ for  other values of $k$. Interestingly we find that for $k=-1$ (as also for log-gas) the distribution converges very fast as can be seen in Figs.~\ref{no-ws-for-gen-k}b and ~\ref{no-ws-for-gen-k}d.
On the other hand for other values of $k$ there is no convergence. In particular for the CM model ($k=2$), our findings~\cite{SM} are thus in disagreement with the generalised version of WS proposed in Ref.~\cite{bogomolny_PRL_2009}. For generic values of $k$, as seen in Fig.~\ref{no-ws-for-gen-k}, the distributions $P_N(s)$ do not show convergence with $N$. Hence, we  look at the distribution of the following natural scaling variable
\beq
\tilde{s}_i  = \frac{\Delta_i -\langle \Delta_i \rangle}{\sigma_{\Delta_i}}.
\label{eqn:tildes}
\eeq
The distribution of this quantity defined as $\tilde{P}_N(\tilde{s})= [1/(N-1)] \sum_{i=1}^{N-1} \langle  \delta (\tilde{s} - \tilde{s}_i)$, is computed numerically for different values of $k$ and  $N$. In Fig.~\ref{tildeP_n(s)} we plot $\tilde{P}_N(\tilde{s})$ for  $k=-1.5,-0.5,0.5$ and $k=1.5$. We find that  $\tilde{P}_N(\tilde{s})$ tends to a  Gaussian form with zero mean and unit variance in the limit $N \to \infty$, except in the range $-1 \leq k \leq 1$. Interestingly, in the range $-1<k<0$, we do not see convergence with $N$ (Fig.~\ref{tildeP_n(s)}b). 
 In the range $0<k<1$  relative fluctuations die out with $N$ in which case one might expect a Gaussian scaling form. Surprisingly, even though the MH nicely predicts the correct scaling exponent $b_k$  the scaling form of the distribution is non-Gaussian (Fig.~\ref{tildeP_n(s)}c). We now present the theoretical arguments which support  the conjecture in Eq.~\eqref{eq:bk} --- based on MH and FT calculations.  

\sectionprl{Microscopic Hessian (MH)}
\label{microscope_hess}
 Computing analytically the variance of the gap for generic values of $k$ is hard (except for $k=-1$ and $k \to 0$). Here we use the microscopic Hessian method ~\cite{Fletcher_computer_1970,BROYDEN_IMA_1970} to estimate the variance for large $N$ for all values of $k$. At zero temperature, the system will be in the ground state characterized by the configuration of positions $y_i$ and corresponding gaps $\Delta_i^{\rm GS} = y_{i+1}-y_i $.
Since the system is at low temperature, we expect that the Hessian of the microscopic Hamiltonian Eq.~\eqref{eqn:1} about the ground state would approximately capture the behavior of the flucutations of the gap. 
The joint distribution of fluctuation of gaps $\delta \Delta_i = \Delta_i - \Delta_i^{\rm GS}$ will be of the form 
\begin{equation}
\label{prob_hess}
\mathcal{P}_{\rm MH}(\{ \Delta_i \}) \sim e^{-\frac{\beta}{2} \sum_{i,j=1}^{N} H_{ij} \delta \Delta_i \delta \Delta_j },
\end{equation}
where the Hessian of the system about the ground state is $H_{ij} =\Big [  \frac{\partial^2{E}}{\partial{\Delta_i}\partial{\Delta_j}}\Big]_{\rm GS}$~\cite{SM}.
The variance of the gap $\sigma^2_{\Delta_i} = {(\beta H)}^{-1}_{ii}$ can thus be obtained by inverting the matrix $H$ numerically. As seen in Fig.~\ref{fig:akbk_k}, the exponent $b_k$ calculated using MH theory matches with the MC result (Eq.~\ref{eq:bk}) except in the regime $-1<k<0$. This is perhaps not surprising since our conjecture suggests that in this regime, the relative fluctuation of the gap, $ { \sigma_{\Delta_i}}/{\langle \Delta_i \rangle} \sim N^{(a_k-b_k/2)}$ does not decrease with system size --- in fact over the range of $N$ considered we see  them increasing (see inset of Fig.~\ref{fig:akbk_k}). Next, we discuss the FT calculation. 

\sectionprl{Field Theory (FT)}   
 As discussed in Ref.~\cite{sanaa2019} the Reisz gas for large-$N$ can be described by a free energy functional $\Sigma [\rho_N] =\mathcal{E}[\rho_N] - \beta^{-1} S[\rho_N]$ corresponding to a macroscopic density profile $\rho_N(x)$ where $\mathcal{E}[\rho_N]$ is the energy and $S[\rho_N] = -N\int  dx \rho_N \log(\rho_N)$ is the entropy functional. The form of the energy functional depends on $k$, being local for $k \geq 1$ and nonlocal for $-2 < k <1$~\cite{sanaa2019, SM}. We use this action to compute the fluctuations of the bulk gap. The probability of a density profile $\rho_N$ is  \cite{Avanish_PRE_2020}
	\begin{equation}
	\label{prob_rho}
	\mathbb{P}[\rho_N] \sim e^{-\beta \delta \Sigma},~\text{with}~\delta \Sigma= \Sigma[\rho_N]-\Sigma[\rho_N^{\rm (eq)}],
	\end{equation}
where $\rho_N^{\rm (eq)}$ is mean thermal density.	For a given  macroscopic density profile $\rho_N(x)$, the gap between two consecutive particles at position $x$ is 
	$\bar{\Delta} = [{N \rho_N(x)}]^{-1}.$	Note that this definition of the gap is different from the gap $\Delta$ defined earlier [above Eq.~\eqref{sdist}] from the microscopic position configuration.	The gap $\bar{\Delta}$ is a coarse grained version of $\Delta$ averaged over many  microscopic  configurations consistent with the macroscopic density $\rho_N(x)$. 
	As the density profile $\rho_N(x)$ fluctuates, the separation $\bar{\Delta}$ also fluctuates. We expect that for large $N $, the fluctuation of $\bar{\Delta}$ and $\Delta$ would have the same  scaling with respect to $N$.
	
	We first find the distribution of the fluctuation $\delta \rho_N(x)$ around the equilibrium profile $\rho_N^{\rm (eq)}(x)$. Writing $\rho_N(x)=\rho_N^{\rm (eq)}(x) +\delta \rho(x)$ in the expression of the action $\delta \Sigma[\rho_N]$ in Eq.~\eqref{prob_rho} and expanding to quadratic order in $\delta \rho(x)$ we get the distribution of the fluctuation profile $\delta \rho(x)$ 
	(see Ref.~\cite{SM}). Note the action $\delta \Sigma$ now becomes an explicit functional of $\delta \rho(x)$ and $\rho_N^{\rm(eq)}(x)$.
	The probability distribution of the fluctuation $\delta \bar{\Delta}$ of the gap, defined as $\bar{\Delta}=\langle \bar{\Delta} \rangle + \delta \bar{\Delta}$, is obtained by using the relation 
\begin{align}
		\delta \bar{\Delta} \approx -\frac{\delta \rho(x)}{N \left(\rho_N^{\rm (eq)}(x)\right)^2},
	\end{align}
	which can be obtained from $\bar{\Delta} = [N \rho_N(x)]^{-1}$. 
	
	For $N$ (large but finite) particles there are $(N-1)$ number of gap variables. In order to find the joint distribution of these (discrete) gap variables from the field theory description, we need to discretise  $\delta \Sigma[\delta \rho, \rho_N^{\rm (eq)}]$. 
	To do so, we discretise the integral in the action $\delta \Sigma$ along the equilibrium positions $\{y_i\}$~\cite{SM}. Recall that the microscopic Hessian was computed about this position configuration in Eq.~\eqref{prob_hess} earlier. Note that $\{y_i\}$, also the minimum energy configuration, leads to the equilibrium macroscopic density $\rho_N^{\rm (eq)}(x)$ which corresponds to mean gaps $\langle \bar \Delta_i \rangle = 1/N\rho_N^{\rm (eq)}(y_i)$. Also note that for large-$N$, $\langle \bar{\Delta}_i \rangle  \approx \Delta_i^{\rm GS}$. We emphasize that this discretisation of the density profile is different from the original microscopic position description of the system.

	We replace the integrals in the expression of $\delta \Sigma$ as $\int_{-l_N}^{l_N} dx \rightarrow \sum_{i} [1/N \rho^{\rm (eq)}_N(y_i)]$ and evaluate the integrand at points $\{y_i\}$.  After some simplifications we get the following joint distribution of the gap variables $\{\delta \bar{\Delta}_i\}$ to leading order in $N$ (see Ref.~\cite{SM} for details): 
		\begin{align}\label{quad_prob_rho}
		&\mathcal{P}_{\rm FT}\left (\{ \delta \bar{\Delta}_i\} \right ) \sim e^{- \frac{\beta}{2} \sum_{i,j=1}^{N}  M_{ij} \delta \bar{\Delta}_i \delta \bar{\Delta}_j}, \text{where}\\
		&M_{ii} = \begin{cases}
			J \zeta(k) k(k+1) N^{k+2} [\rho^{\rm (eq)}_N(y_i) ]^{k+2} &\text{for }   k>1 \\
			2 J N^{k+2} [\rho^{\rm (eq)}_N(y_i) ]^{k+2} &\text{for }  0<k<1 \\
			N^2 \beta^{-1} [\rho^{\rm (eq)}_N(y_i)]^2 & \text{for }   -2<k <0,			
		\end{cases}
		\notag \\ 
		&M_{ij } = \begin{cases}
		0 &\text{for } k>1,\\
		J N^2 ~\text{sgn}(k) \frac{ \rho^{\rm (eq)}_N(y_i) \rho^{\rm (eq)}_N(y_j)}{|y_i - y_j|^k} &\text{for } -2<k<1. 
		\end{cases}	\label{M_ij}
	\end{align}
 For the diagonal term, it is interesting to note~\cite{SM} that, for $-2<k<0$, the contribution from entropy is dominant whereas, for $k>0$, the contribution from energy is dominant.
The variance of $\bar{\Delta}$ is given by $\langle \delta \bar{\Delta}_i^2 \rangle= (\beta M)^{-1}_{ii}$.  Assuming that the inverse of the dominant term of the matrix $M$ (see Eqs.~(28,29) in Ref.~\cite{SM}) dictates the scaling of the variance we arrive at the conjecture in Eq.~\eqref{eq:bk}. 
We also compute the variance from a direct numerical inversion of the matrix $M$ and  as seen in Fig.~\ref{fig:akbk_k} we find very good agreement with the conjecture in Eq.~\eqref{eq:bk} for all $k$ values. The deviation from the MC results  are possibly due to statistical errors, slow equilibration and finite-size effects.

\sectionprl{Conclusions} In this Letter, we have studied the nearest neighbour gap statistics for a harmonically confined Riesz gas, in particular the variance and the distribution.  The variance of the bulk gap is characterized by the exponent $b_k$ for which we conjecture a form, Eq.~\eqref{eq:bk}, for the $k$-dependence. We provided support for this through direct MC simulations, and numerics based on small fluctuations theories such as microscopic Hessian and quadratic field theory.  We studied the normalized gap distribution, $P_N(s)$ and find a convergence, with $N$, for $k=0^+,-1$. For other values of $k$, $P_N(s)$ does not converge with increasing $N$.  This leads us to study $\tilde{s}_i$  [gap normalized by  fluctuations, see Eq.~\eqref{eqn:tildes}]. As summarized in Fig.~\ref{fig:summary}, for $-2<k<-1$ and $k>1$ we found that the scaling form of $\tilde{P}_N(\tilde{s})$ is Gaussian while for all other $k$ values, we find strong non-Gaussian behaviour. In fact, for $-1<k<0$, we found that there is no convergence with $N$. 
Moreover in this regime, the  fluctuations are of the same order as the mean, leading to the failure of the Hessian theory. Remarkably, the quadratic field theory approach is able to predict the expected scaling exponent even in this regime. It is worth re-emphasizing that the analytical microscopic treatment of fluctuations is extremely difficult. We have proposed two different analytical approaches which are able to successfully capture the main features seen by direct simulations: (i) mapping between the microscopic variables and the coarse-grained macroscopic density field. This provides an enormous simplification for the otherwise intractable and highly non-local microscopic model. (ii) Hessian approximation which results in an all-to-all connected Harmonic network and provides a powerful tool for tackling long-ranged systems.
 Some interesting outstanding problems include understanding of the non-Gaussian behaviour, including large deviations, of the gap distribution and its analytical derivation for special cases such as the 1dOCP ($k = -1$), CM ($k=2$) and hard rods ($k\to \infty$).

\sectionprl{Acknowledgements} 
We thank A. Flack for her simulation results for the gap distribution in the bulk for $k=-1$. We thank S. N. Majumdar and G. Schehr for very useful discussions and a careful reading of the manuscript. MK would like to acknowledge support from the project 6004-1 of the Indo-French Centre for the Promotion of Advanced Research (IFCPAR), Ramanujan Fellowship (SB/S2/RJN-114/2016), SERB Early Career Research Award (ECR/2018/002085) and SERB Matrics Grant (MTR/2019/001101) from the Science and Engineering Research Board (SERB), Department of Science and Technology, Government of India. This research was supported in part by the International Centre for Theoretical Sciences (ICTS) for enabling discussions during the program - Fluctuations in Nonequilibrium Systems: Theory and applications (Code:ICTS/Prog-fnsta2020/03). SS, JK, AD, MK and AK acknowledge support of the Department of Atomic Energy, Government of India, under Project No. RTI4001. 

\bibliography{Main_draft.bib}

\end{document}


	
	
	\date{\today}
	\newcommand{\titlename}{Supplementary material for `Gap statistics for confined particles with power-law interactions'}

	\title{\titlename}
	
\author{S. Santra}
\email{saikat.santra@icts.res.in} 
\address{International Centre for Theoretical Sciences, Tata Institute of Fundamental Research, Bengaluru -- 560089, India}

\author{J. Kethepalli}
\email{jitendra.kethepalli@icts.res.in}
\address{International Centre for Theoretical Sciences, Tata Institute of Fundamental Research, Bengaluru -- 560089, India}

\author{S. Agarwal}
\email{sanaa.agarwal@colorado.edu} 
\address{Department of Physics, University of Colorado, Boulder, Colorado 80309, USA}
\author{ A. Dhar}
\email{abhishek.dhar@icts.res.in}
\address{International Centre for Theoretical Sciences, Tata Institute of Fundamental Research, Bengaluru -- 560089, India}

\author{M. Kulkarni}
\email{manas.kulkarni@icts.res.in}
\address{International Centre for Theoretical Sciences, Tata Institute of Fundamental Research, Bengaluru -- 560089, India}

\author{A. Kundu}
\email{anupam.kundu@icts.res.in}	
\address{International Centre for Theoretical Sciences, Tata Institute of Fundamental Research, Bengaluru -- 560089, India}

	\maketitle
	
	\tableofcontents 
	
	\vspace{0.8cm}
	This supplementary is organized as follows. In Sec.~\ref{sec:model}, we give the details of the model and recap some relevant previous results. In Sec.~\ref{mean_and_variance} we give details of the computation of the variance of the bulk gap using Monte-Carlo simulation (MC), microscopic Hessian (MH) and quadratic field theory (FT). In Sec.~\ref{Interesting} some interesting features of gap statistics for $-1\leq k \leq 0$ are discussed. In Sec.~\ref{sec:CM_supp}, we study the gap distribution for the Calogero-Moser (CM) model ($k=2$) and discuss reasons for disagreement with previous studies~\cite{bogomolny_PRL_2009}.

	\section{Model: Riesz gas and density profile}
	\label{sec:model}
	In the main text we considered a classical system of $N$ particles on a line confined by a harmonic potential and interacting with each other via a repulsive power law interaction of the form $\text{sgn}(k)|r|^{-k}$, where $r$ is the distance between two particles. The potential energy of the system is given by ($ \forall k > -2$)~\cite{riesz_ASU_1938} \\
	\bea
	E \left(\{x_i\} \right )=\sum_{i=1}^{N} \frac{x^2_i}{2}+ \frac{J ~\text{sgn}(k)}{2} \sum_{i\neq j} \frac{1}{\mid x_i-x_j \mid ^k}, 
	\label{eqn:1}
	\eea
	where $J>0$ and  $\text{sgn}(k)$ ensures a repulsive interaction. Using large $N$ field theory, the  average thermal density is calculated recently~\cite{sanaa2019} $\forall$ $k>-2$. The average density, $ \rho^{\rm (eq)}_N $, for large $N$ is independent of the inverse temperature, $\beta$ and is described by the following scaling form
	\beq
	\rho^{\rm (eq)}_N(x)   \equiv \frac{1}{N} \sum_{j=1}^{N} \langle \delta (x-x_j)\rangle = \frac{1}{l_k N^{\alpha_k}} F_k \big(\frac{x}{l_k N^{\alpha_k}}\big),
	\eeq
	where $\langle...\rangle$ denotes an equilibrium average and the exponent $\alpha_k$ is given by 
	\begin{align}
	\alpha_k  =	\begin{cases} \frac{k}{k+2} ~\rm{for}~ k>1 \\
	\frac{1}{k+2} ~\rm{for}~  -2<k<1.
	\end{cases}	
	\end{align}
	The scaling function $F_k(y)$, supported over $y \in [-1/2,1/2]$, is given explicitly by 
	\beq
	F_k(y)=\frac{1}{B\big(\gamma_k+1,\gamma_k+1\big)} \big(\frac{1}{4}-y^2\big)^{\gamma_k},
	\eeq 
	where the exponent $\gamma_k$ is given by 
	\begin{align}
	\gamma_k = \begin{cases}   \frac{1}{k} ~~~~\rm{for}~ k>1 \\
	\frac{k+1}{2} ~\rm{for}~ -2<k<1,
				\end{cases}
	\end{align}
	and the system size independent length scale is explicitly 
	\begin{align}
	l_k = \begin{cases} \Big (\frac{\big (2 J \zeta(k) (k+1) \big )^{1/k}}{B(1+1/k,1+1/k)} \Big )^{\frac{k}{k+2}} ~~~~~~~~~\rm{for}~  k>1 \\
	\Big (	\frac{J|k| \pi (k+1)}{\sin \big [\frac{\pi}{2}(k+1)\big] B(\frac{k+3}{2},\frac{k+3}{2})}\Big )^{\frac{1}{k+2}} ~\rm{for}~   -2<k<1, \end{cases}		
	\end{align}	
	where $B$ is beta function. We use these densities to study the statistical properties of interparticle bulk gaps in detail.

	
	
	\section{Variance of the bulk gap}
	\label{mean_and_variance}
	In this section we discuss the calculation of the variance of bulk gap  $\Delta_i =x_{i+1}-x_i$  ($ 1 \ll i \ll N-1$), where $\{x_i\}$ are the positions of the particles. As discussed in the main text, the mean of the bulk gap scales with system size as $\langle \Delta_i \rangle \sim N^{\alpha_k}/N =N^{-a_k}$ where $a_k=1-\alpha_k$, i.e., 
	\bea
	\label{eq:ak}
	a_k=	 \begin{cases} \frac{2}{k+2} & \text{for } k>1 \\
	\frac{k+1}{k+2} &\text{for}   -2<k<1. \end{cases}	
	\eea
	Our conjectured form for  the variance of bulk gap  is $\sigma^2_{\Delta_i}\sim N^{-b_k}$, with
	\begin{align}
	b_k  = \begin{cases} 2 ~~ &{\rm{for}}~~ k>1 \\
	1+k & {\rm for}  ~ 0<k<1 \\
	2(k+1)/(k+2) & {\rm for}  ~ -1<k<0  \\   
	1+k  ~~   &{\rm for}  ~~ -2<k<- 1. \end{cases}
	\label{eq:bk}
	\end{align}
	Here we present the details of the microscopic Hessian (MH) and quadratic field theory (FT) which are used to compute the variance and provide support for the above conjectured form. 
	\subsection{Low temperature dynamics}
	\begin{figure}[t]
		\begin{minipage}[H]{0.49\textwidth}
			\centering		\includegraphics[width=\textwidth]{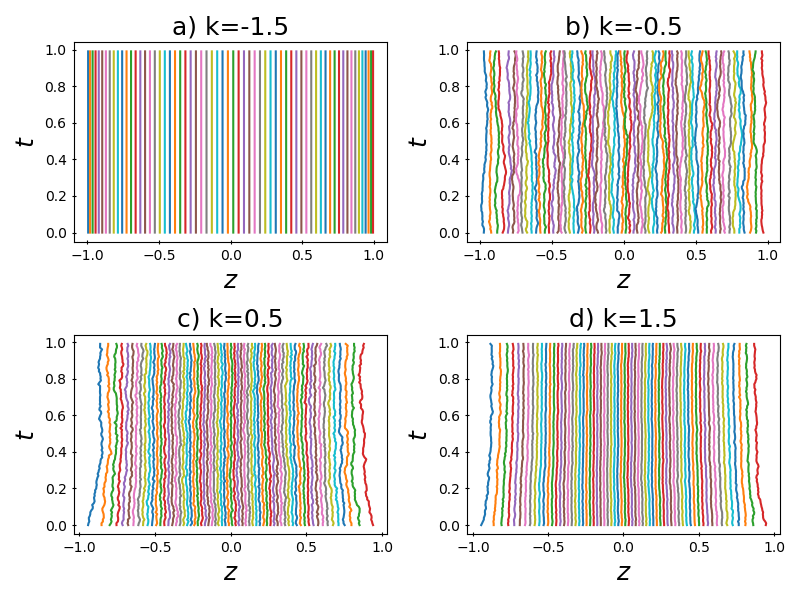}		
		\end{minipage}
		\caption{Spacetime trajectories of particles evolving  under overdamped Langevin equations and starting from their equilibrium positions $\{\bar{y}_i\}$. The  x-axis represents the scaled position $z=x/(l_k N^{\alpha_k})$ and we show results for different values of $k$. We took $\gamma=1, \beta=16$ and $N=64$.}
		\label{fig:dynamics}
	\end{figure}
	
	To get some intuition about the microscopic fluctuations of particle positions in our many-body system, it is instructive to  study the overdamped Langevin dynamics of the particles at low temperatures ($T=\beta^{-1} < N^{2\alpha_k}$). We thus consider the equations of motion
	\beq
	\dot{x_i}(t) = -{\partial_{x_i}{E}}/{\gamma} + \sqrt{(2/\gamma \beta)} \eta_i (t),
	\eeq
	where $\eta_i(t)$ are Gaussian white noise terms with zero mean and unit variance.  The initial condition are taken to be $x_i(0)=\bar{y}_i$, where $\bar{y}_i$ are the equilibrium positions. This is computed from the equilibrium density profile using 
	\begin{equation}
	\label{eq:wibar}
	i=N\int_{-l_k}^{\bar{w}_i} \rho^{\rm (eq)}_N(x) dx,  
	\end{equation}
	and then taking the symmetrised form 
	\begin{equation}
	\label{eq:yibar}
	\bar{y}_i = \frac{\bar{w}_i-\bar{w}_{N-i+1}}{2}.
	\end{equation}
	From the spacetime trajectories shown in Fig.~\ref{fig:dynamics} we see that for all $k>-2$, the particles fluctuate around their equilibrium positions.  We find that the fluctuations about the equilibrium configurations are large for $-1<k<0$ compared to other values of $k$ --- this is consistent with our findings on the  gap flutuations relative to mean (see Sec.~\ref{Interesting}).
	\subsection{Details of Monte-Carlo (MC) simulations}
	We compute the mean and variance of the bulk gap using Monte-Carlo simulations. For a given processor, we disregard about initial $10^6$ MC cycles. By MC cycle, we mean $O(N)$ Metropolis steps. We then collect data of about $10^6$ samples. This process is simultaneously executed in around $N/2$ processors thereby making number of samples to be around $10^8$. Needless to mention, the exact numbers depend on whether we achieve satisfactory convergence of our results for a given $N$ and $k$. For all the spacing distributions (which contain a total of $N-1$ gaps ), we have used $10^6$ MC samples.  Both  $J_0$ and $J$ are taken to be unity for log-gas ($k=0^+$) and for any other values of $k>-2$ respectively.  
	
	\subsection{Microscopic Hessian}
	At equilibrium the particles are sampled from the Gibbs Boltzmann distribution 
	\beq
	P_G(\{x_i\}) = \frac{e^{-\beta E\{x_i\}}}{Z},
	\eeq
	where the partition function $Z=\int \prod_i dx_i e^{-\beta E\{x_i\}}$. Then the joint distribution of all gaps is given by
	\beq
	\label{dist:Delta_i}
	\mathcal{P}(\{\Delta_i\}) = \int \left( \prod_{i=1}^{N-1} dx_i \delta \left(\Delta_i -\left(x_{i+1}-x_i\right)\right) \right) e^{-\beta E(\{x_i\})}.
	\eeq
	Using this distribution, in principle one can calculate mean and the variance of the $i^{\rm th}$ gap. However the integrals over microscopic positions are difficult to compute, so we only solve them approximately. At zero temperature,
	the system will be in the ground state characterised by
	the configuration of positions $y_i$ and corresponding gaps $\Delta_i^{\rm GS}= y_{i+1} - y_i$ . At low temperatures we expect that the Hessian of the microscopic Hamiltonian Eq.~\eqref{eqn:1} about the ground state would approximately capture the behavior of the flucutations of the gap. The positions  $y_i$ are obtained by minimising the energy using the Broyden-Fletcher-Goldfarb-Shanno (BFGS) algorithm ~\cite{Fletcher_computer_1970,BROYDEN_IMA_1970}. Under the Hessian approximation the joint distribution of fluctuation of gaps $\delta \Delta_i = \Delta_i - \Delta^{\rm GS}_i$
	\begin{equation}
	\label{prob_hess}
	\mathcal{P}_{MH}(\{\Delta_i \}) \sim e^{-\beta \left [ \frac{1}{2} \sum_{i,j=1}^{N} H_{ij} \delta \Delta_i \delta \Delta_j \right ]},
	\end{equation}
	where the Hessian of the system about the equilibrium configuration is $H_{ij} =\Big [  \frac{\partial^2{E}}{\partial{\Delta_i}\partial{\Delta_j}}\Big]_{\rm GS}$.
	The matrix $H_{ij}$ is obtained as follows. The Hessian $\tilde{H}_{ij}$ of the Hamiltonian Eq.~\eqref{eqn:1} in terms of the fluctuation of positions around their equilibrium values is given by
	\begin{equation}
	\begin{split}
	\tilde{H}_{ij}&=\Big [  \frac{\partial^2{E(\{x \})}}{\partial{x_i}\partial{x_j}}\Big]_{\rm GS}\\&=\delta_{ij} \Big [ 1+ \sum_{n \neq i}^{N} \frac{J \text{sgn}(k)k(k+1)}{ \big (y_i-y_n \big )^{k+2}}\Big ] \\& \quad\quad-(1-\delta_{ij})\frac{J \text{sgn}(k) k(k+1)}{ \big (y_i-y_j \big )^{k+2}}.
	\end{split}
	\end{equation}
	%
	To get the Hessian $H_{ij}$ we change position to the gap variables, $\Delta_i=x_{i+1}-x_i$ for $i=1,\ldots,N-1$ and the centre of mass coordinate which we denote by $\Delta_N  \equiv \sum_{i=1}^{N}x_i/N$ by the transformation
	\begin{align}
	\label{posn_gap}
	x_j & =\Delta_N -\sum_{i=j}^{N-1} \Delta_i + \sum_{i=1}^{N-1} \frac{i}{N} \Delta_i \\
	& = \sum_{j=1}^{N} A_{ij} \Delta_j,
	\end{align}
	where $A$ is a matrix of dimension $N \times N$ with matrix elements
	\bea
	A_{ij} =
	\begin{cases} \frac{j}{N} &\text{for }   i>j \\  (\frac{j}{N}-1)  & \text{for } i \leq j(\neq N) \\ 
		1 & \text{for  } j=N,i=1,...,N.
	\end{cases}	
	\eea
	The quadratic Hamiltonian in terms of new variables becomes $ \frac{1}{2} \sum_{i,j=1}^{N} H_{ij}\delta \Delta_i \delta \Delta_j$,
	where $H = A^{T} \tilde{H} A$. Using Eq.~\eqref{prob_hess} we compute the variance of bulk gap from the relation
	$\sigma^2_i = \langle \delta \Delta^2_i \rangle ={(\beta H)^{-1}_{ii}},$
	where $H^{-1}_{ii}$ is found numerically.

	\subsection{Quadratic field theory}
	As discussed in Ref.~\cite{sanaa2019} the Reisz gas for large-$N$ can be described by a free energy functional $\Sigma [\rho_N] =\mathcal{E}[\rho_N] - \beta^{-1} S[\rho_N]$ corresponding to a macroscopic density profile $\rho_N(x)$ where $\mathcal{E}[\rho_N]$ is the energy and 
	\bea
	S[\rho_N] = -N\int_{-\infty}^{\infty}  dx \rho_N \log(\rho_N)
	\eea
	is the entropy functional. The form of the energy functional depends on $k$, being local for $k \geq 1$ and non-local for $-2 < k <1$~\cite{sanaa2019}. 
	The energy functional,$\mathcal{E}[\rho_N(x)]$ (\cite{sanaa2019})  is given by
	\begin{align}
		\mathcal{E}[\rho_N(x)] \approx  
		&\frac{N}{2} \int_{-\infty}^{\infty} x^2 \rho_N(x) dx -\mu \left (\int_{-\infty}^{\infty} \rho_N(x) d x-1 \right ) \notag \\ 
		&+ J \zeta(k) N^{k+1} \int_{-\infty}^{\infty} [\rho_N(x)]^{k+1} dx,
		\label{local}
	\end{align}
	for $k>1$ whereas for $-2<k<1$, it is given by
		\begin{align}\label{field:kg1}
			&\mathcal{E}[\rho_N(x)] \approx 
			\frac{N}{2} \int_{-\infty}^{\infty} x^2 \rho_N(x) dx -\mu \left (\int_{-\infty}^{\infty} \rho_N(x) d x-1 \right )\notag \\ 
			&+ \frac{J\text{sgn}(k) N^2}{2}  PV \int_{-\infty}^{\infty} dx \int_{-\infty}^{\infty} dy \frac{\rho_N(x) \rho_N(y)}{|x-y|^k}.			
		\end{align}  
	where $PV$ stands for principal value. As our aim is to find joint distribution of $(N-1)$ gaps, it will be useful to approximate the principal value integral in the Eq.~\eqref{field:kg1}. To do so we assume that there is a particle at $x$ and break up the $y$ integrals in the three regions, ($-\infty,x-\bar{\Delta}$),($x-\bar{\Delta},x+\bar{\Delta}$) and ($x+\bar{\Delta},\infty$), where $\bar{\Delta}$ is  the separation between two adjacent particles. Noting that the $\rho_N(y)$ vanishes in the window ($x-\bar{\Delta},x+\bar{\Delta}$), we rewrite Eq.~\eqref{field:kg1} as
\begin{align}\label{field:kg2}
			&\mathcal{E}[\rho_N(x)] \approx 
			\frac{N}{2} \int_{-\infty}^{\infty} x^2 \rho_N(x) dx -\mu \left (\int_{-\infty}^{\infty} \rho_N(x) d x-1 \right )\notag \\ 
			&+ \frac{J\text{sgn}(k) N^2}{2}   \int_{-\infty}^{\infty} dx \Bigg [ \int_{-\infty}^{x-\bar{\Delta}}+ \int_{x+\bar{\Delta}}^{\infty} \Bigg ]   dy \frac{\rho_N(x) \rho_N(y)}{|x-y|^k}. 
		\end{align}	
	\begin{figure}[t]
		\centering		
		\includegraphics[width=0.49\textwidth]{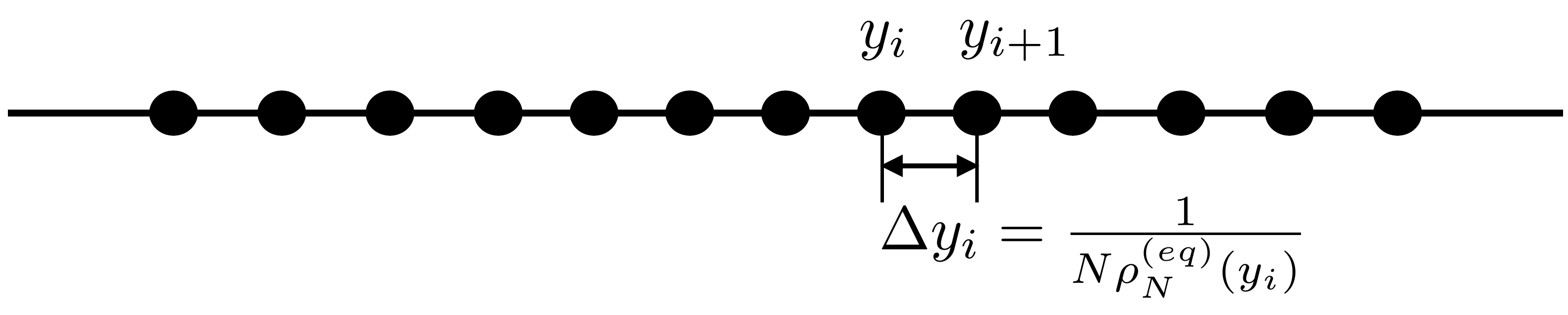}		
		\caption{Schematic description of the discretisation of the integral in the action $\delta \Sigma$. }
		\label{discrtsn}
	\end{figure}
	 The separation $\bar{\Delta}$ at $x$ can be expressed in terms of the density as 
	\beq
	\label{Delta-rho-rela}
	\bar{\Delta}= \frac{1}{N \rho_N(x)}.
	\eeq
	\begin{figure}[h]
		\begin{minipage}[H]{0.45\textwidth}
			\centering		\includegraphics[width=\textwidth]{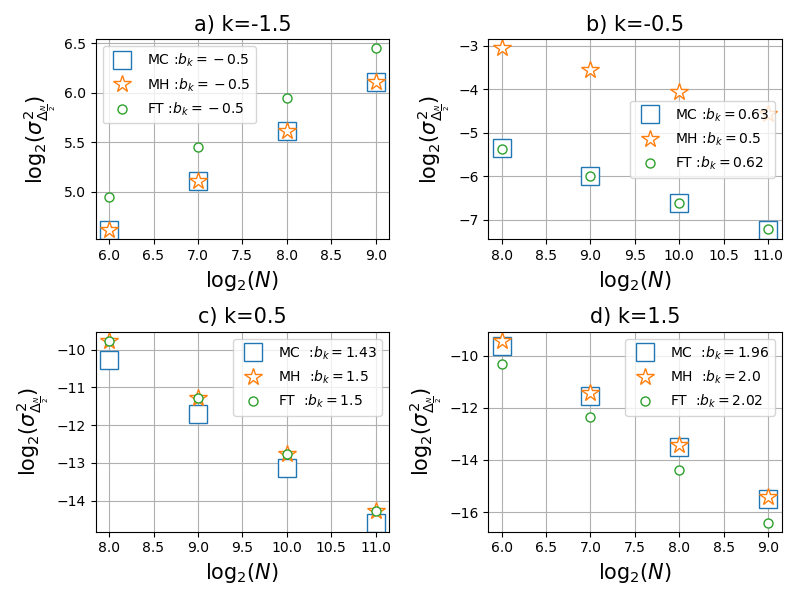}		
		\end{minipage}
		\caption{Comparison of actual values of the variance of mid-gap calculated using all three techniques, MC, MH and FT for $\beta=1$. }
		\label{comparison}
	\end{figure}		
	As the density fluctuates, the gap profile also fluctuates and we are interested to compute the variance of the gap in the bulk of the system. The average value of the gap is given by the equilibrium density profile $\rho_N^{\rm (eq)}(x)$ {\it i.e.} 
	$\langle \bar{\Delta} \rangle= \frac{1}{N \rho_N^{\rm (eq)}(x)}$.
	The probability of a given density profile $\rho_N(x)$ \cite{Avanish_PRE_2020} is 
	\begin{equation}
	\label{eq:prob_rho}
	\mathbb{P}[\rho_N] \sim e^{-\beta \delta \Sigma},~\text{with,}~\delta \Sigma= \Sigma[\rho_N]-\Sigma[\rho_N^{\rm (eq)}]
	\end{equation} 
	where $\Sigma[\rho_N] = \mathcal{E}[\rho_N(x)] - \beta^{-1}S[\rho_N(x)]$
	is the free energy functional.
	Writing the fluctuations of density profile about the equilibrium as  $\rho_N(x)=\rho^{\rm (eq)}_N(x) +\delta \rho(x)$ and using the relation in Eq.~\eqref{Delta-rho-rela} we obtain
	\begin{equation}\label{bar_rho}
		\bar{\Delta}(x) = \langle \bar{\Delta} \rangle + \delta \Delta= \frac{1}{N \rho_N^{\rm (eq)}(x)} - \frac{\delta \rho(x)}{N \left(\rho_N^{\rm (eq)}(x)\right)^2}.
	\end{equation}
	Using this Eq.~\eqref{bar_rho} we expand the exponent $\delta \Sigma= \Sigma[\rho_N]-\Sigma[\rho_N^{\rm (eq)}]$ in Eq.~\eqref{eq:prob_rho} in powers of $\delta \rho(x)$ (up to second order in $\delta \rho$) to get 
	
	\begin{align}
	\begin{split}
	\delta \Sigma [\delta \rho,\rho_N^{\rm (eq)}]& \approx  \frac{J\zeta(k)k(k+1)N^{k+1}}{2} \\
	& \times  \int_{-l_k}^{l_k}dx~[\rho_N^{\rm (eq)}]^{k-1}~\delta \rho(x)^2 \\
	&
	+ \frac{N}{2 \beta}  \int_{-l_k}^{l_k}dx~\frac{\delta \rho(x)^2}{\rho_N^{\rm (eq)}(x)}, ~\quad\text{for}~k>1, 
	\end{split} \label{dSigma-k>1}
	\end{align}
	and
	\begin{align}	
	\begin{split}
	&\delta \Sigma [\delta \rho,\rho_N^{\rm (eq)}] \\
	&\approx
	\frac{2 J\text{sgn}(k)N^{k+1}}{2}\int_{-l_k}^{l_k}dx~[\rho_N^{\rm (eq)}]^{k-1}~\delta \rho(x)^2 \\
	&+\frac{J\text{sgn}(k)N^{2}}{2}\int_{-l_k}^{l_k}dx \left [ \int_{-l_k}^{x-\langle \bar{\Delta} \rangle}+\int^{l_k}_{x+\langle \bar{\Delta} \rangle}\right] dy \\
	&~~~~~~~
	\times~\frac{\delta \rho(x)~\delta \rho(y) }{|x-y|^k},\\ 
	&~~
	+ \frac{N}{2 \beta}\int_{-l_k}^{l_k}dx~\frac{\delta \rho(x)^2}{\rho_N^{\rm (eq)}(x)}, \quad \text{for}~-2<k<1.
	\end{split} \label{dSigma-k<1}
	\end{align}
It is pertinent to note that, in the quadratic approximation of the field theory, the local term appears due to the fluctations of the coarse grained gap $\bar{\Delta}$ in the non-local term of the energy $\mathcal{E}[\rho_N(x)]$ for $-2<k<1$. However for $k>1$ the field theory in Eq.~\eqref{local} is local to leading order in $N$. This leads to Eq.~\eqref{dSigma-k>1} which further implies $\sigma^2_{\Delta_{N/2}} \sim N^{-2}$. We have neglected the contribution from non-local terms in the field theory because they are expected to contribute to $\sigma^2_{\Delta_{N/2}}$ at most at the same order ($N^{-2}$) or less, thereby leaving the scaling exponent value $b_k$ unchanged.    
For $N$ (large but finite) particles there are $(N-1)$ number of gap variables. In order to find the joint distribution of these (discrete) gap variables from the field theory description, we need to discretise $\delta \Sigma$ given in Eqs.~\eqref{dSigma-k>1} and \eqref{dSigma-k<1}. To do so, we discretise the integral in the action $\delta \Sigma$ {about} the equilibrium positions $\{y_i\}$ (see Fig.~\ref{discrtsn}). Recall that the microscopic Hessian was computed about this position configuration in Eq.~(7) of the main text. Note that the equilibrium (in this case the minimum energy) position configuration $\{y_i\}$ correspond to equilibrium macroscopic density $\rho_N^{\rm (eq)}(x)$. Hence replacing the integrals as $\int_{-l_k}^{l_k} dx \rightarrow \sum_{i} \frac{1}{N \rho^{\rm (eq)}_N(y_i) }$ and performing some simplifications we get the following joint distribution of the gap variables $\{\delta \bar{\Delta}_i\}$: 
		\begin{align}\label{quad_prob_rho-sm}
		&\mathcal{P}_{\rm FT}\left (\{ \delta \bar{\Delta}_i\} \right ) \sim e^{- \frac{\beta}{2} \sum_{i,j=1}^{N}  M_{ij} \delta \bar{\Delta}_i \delta \bar{\Delta}_j}, \text{where}\\
		&M_{ii} = \begin{cases}
		J \zeta(k) k(k+1) N^{k+2} [\rho^{\rm (eq)}_N(y_i) ]^{k+2} &\text{for }   k>1 \\
		2 J N^{k+2} [\rho^{\rm (eq)}_N(y_i) ]^{k+2} &\text{for }  0<k<1 \\
		N^2 \beta^{-1} [\rho^{\rm (eq)}_N(y_i)]^2  &\text{for }   -2< k <0,		
		\end{cases}
		\notag \\
		&	 M_{ij } = \begin{cases}
		0 &\text{for }   k>1\\		
		J N^2 ~\text{sgn}(k) \frac{ \rho^{\rm (eq)}_N(y_i) \rho^{\rm (eq)}_N(y_j)}{|y_i - y_j|^k} &\text{for } -2<k<1,
		\end{cases}
		\end{align}	 
	in the leading orders in $N$. Using the scaling form  $\rho^{\rm (eq)}_N(x) = \frac{1}{l_k N^{\alpha_k}} F_k \left(\frac{x}{l_k N^{\alpha_k}}\right)$ with $\alpha_k =\frac{k}{k+2}$ for $k>1$ and  $\frac{1}{k+2}$ for $-2<k<1$, we find the following 
	system size dependence of the matrix elements (see the next subsection for details):
		\bea
		\label{eq:mij}
		M_{ii}& \sim &\begin{cases} 
			\mathcal{O} \left (N^2 \right ) &\text{for }   k>1 \\
			\mathcal{O} \left (N^{k+1} \right )  & \text{for } 0<k<1 \\
			\mathcal{O} \left (N^{2(k+1)/(k+2)} \right ) 	& \text{for } -2<k<0, 
		\end{cases} 	\\
		M_{ij} & \sim &	 \begin{cases}
			  {0} &\text{for }   k>1  \\ 
			\mathcal{O} \left (N^{k+1} \right )  & \text{for } -2<k<1,			  
		\end{cases}
		\eea 	
	where we have ignored  terms with $|i-j| \sim \mathcal{O}(N)$. 
	From Eq.~\eqref{quad_prob_rho-sm}, it is easy to compute the variance of middle gap $\Delta_{N/2}$ given by $\langle \delta \bar{\Delta}_{N/2}^2 \rangle= { [(\beta M)^{-1}]_{\frac{N}{2} \frac{N}{2}}}$. This can be easily evaluated by numerically inverting the matrix $M$ and we find the following scaling of the variance of middle gap
	\bea
	\sigma^2_{\Delta_{N/2}} \sim  \begin{cases}
		\mathcal{O} \left (N^{-2} \right ) &\text{for }   k>1 \\
		\mathcal{O}\left (N^{-(k+1)}\right)  &\text{for } 0<k<1 \\
		\mathcal{O}\left (N^{-2(k+1)/(k+2)}\right) & \text{for } -1<k<0 \\
		\mathcal{O}\left (N^{-(k+1)} \right )  &\text{for } -2<k<-1,		 
	\end{cases}
	\eea
	which match remarkably well with the large-N scaling obtained from MC simulation. 
	We notice that the inverse of the dominant term in the matrix $M_{ij}$ irrespective of whether it is diagonal or off-diagonal gives  us the  expected conjectured form for the scaling exponent $b_k$.
	\subsection{System size scaling of the matrix elements $M$ and $H$}
	
	
	The variance of the bulk gap and the coarse grained bulk gap are described by  MH and FT respectively and these are in turn governed by the matrices $H$ and $M$ respectively. Since they quantify the similar physical quantities it is natural to compare the matrices  element wise which is shown in Table~\eqref{table:matrices}.
	\begin{table}[h]
		\centering
		\begin{tabular}{| c | c | c | c | c |}
			\hline
			& \multicolumn{2}{c} {\textbf{$M_{ij}$}} & \multicolumn{2}{|c|} {\textbf{$H_{ij}$}}  \\
			\hline
			\textbf{Range of $k$} & $i=j$ & $|i-j| \sim \mathcal{O}(1)$ & $i=j$ & $|i-j| \sim \mathcal{O}(1)$ \\ 
			\hline
			$k>1$ & $N^2$ &  $ {0}$ & $N^2$ &  $N^2$ \\
			\hline  
			$0<k<1$ & $N^{1+k}$ &  $N^{1+k}$ &  $N^{1+k}$ &  $N^{1+k}$  \\
			\hline  
			$-1<k<0$ & $N^{\frac{2(1+k)}{(k+2)}}$ &  $N^{1+k}$ & $N$ &  $N^{1+k}$ \\
			\hline
			$-2<k<-1$ & $N^{\frac{2(1+k)}{(k+2)}}$ &  $N^{1+k}$ & $N$ &  $N^{1+k}$ \\
			\hline
		\end{tabular}
		\caption{System size dependence of the matrix elements $M_{ij}$ and $H_{ij}$ for different values of $i$ and $j$. As we are interested mainly in the bulk part of the system we ignore such terms where $(i-j) \sim \mathcal{O}(N)$.} \label{table:matrices}
	\end{table}
	We notice that the elements of matrices $H$ and $M$ have similar system size scaling for $0<k<1$, while  for $k>1$ off-diagonal terms  and for $-2<k<0$  the diagonal terms differ. 	
	The variance of the bulk gap are found by the numerical inversion of these matrices $H$ and $M$.
	The scaling behaviour of the elements of the inverse matrices are presented in Table.~\eqref{table:Inv}

	\begin{table}[h]
		\resizebox{\columnwidth}{!}{%
			\begin{tabular}{| c | c | c | c | c |}
				\hline
				& \multicolumn{2}{c} {\textbf{$M^{-1}_{ij}$}} & \multicolumn{2}{|c|} {\textbf{$H^{-1}_{ij}$}}  \\
				\hline 
				\textbf{Range of $k$}   & $i=j$ & $|i-j| \sim \mathcal{O}(1)$ & $i=j$ & $|i-j| \sim \mathcal{O}(1)$ \\ 
				\hline
				$k>1$ & $N^{-2}$ &  $ {0}$ & $N^{-2}$ &  $N^{-2}$ \\
				\hline  
				$0<k<1$ & $N^{-(1+k)}$ &  $N^{-(1+k)}$ & $N^{-(1+k)}$ &  $N^{-(1+k)}$ \\
				\hline  
				$-1<k<0$ & $N^{\frac{-2(1+k)}{(k+2)}}$ &  $N^{\frac{-2(1+k)}{(k+2)}}$ & $N^{-(1+k)}$ &  $N^{-(1+k)}$ \\
				\hline
				$-2<k<-1$ & $N^{-(1+k)}$ &  $N^{-(1+k)}$ & $N^{-(1+k)}$ &  $N^{-(1+k)}$ \\
				\hline
			\end{tabular}%
		}
		\caption{System size dependence of the matrix elements $M^{-1}_{ij}$ and $H^{-1}_{ij}$ for different values of $i$ and $j$. Again we ignore such terms where $(i-j) \sim \mathcal{O}(N)$.}
		\label{table:Inv}
	\end{table}
	We observe that the inverse of the dominant term in the matrix $M$ gives the scaling similar to the variance calculated by numerical inversion of these matrices. This leads us to the conjecture for the variance given in Eq.~\eqref{eq:bk}.  It is clear from the Table.~\ref{table:Inv} that except in the regime $-1<k<0$ the microscopic Hessian theory captures the system size dependence of variance of bulk gap whereas the quadratic field theory gives correct exponent $b_k$ (consistent with MC simulation) for all $k>-2$.
	
	The $N$ dependence of the variance as given in Eq.~\eqref{eq:bk} is verified numerically and presented in the Fig.~3 of the main text. However, this does not contain information on the actual values obtained using different methods (MC, MH and FT). This comparison is presented in Fig.~\ref{comparison}. While the slopes are consistent with the conjecture, the actual values do not always match which is not surprising. We find that whenever MH and MC system size scaling agree  [$k \notin (-1,0)$], they even match quantitatively for large-$N$. Remarkably, in the region $-1<k<0$, the MC and FT results match quantitatively as well. 
	

	
	\begin{figure}[t]
		\begin{minipage}[H]{0.48\textwidth}
			\centering		\includegraphics[width=\textwidth]{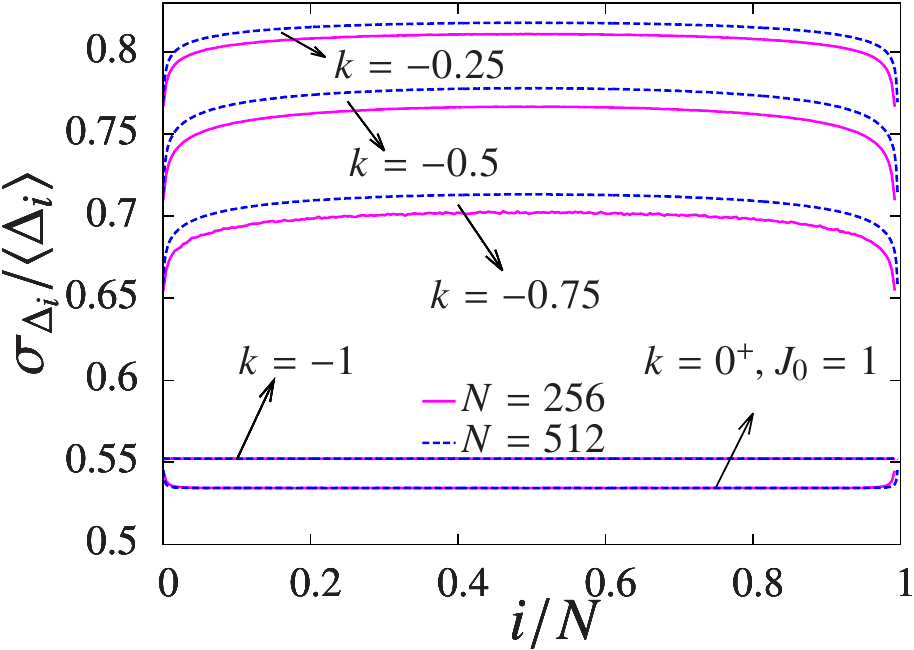}		
		\end{minipage}
		\caption{ Relative fluctuations $\frac{\sigma_{\Delta_i}}{\langle \Delta_i \rangle}$ are shown as a function of $i$ for different system sizes $N$ as well as for different $k$ in the regime $-1 \leq k \leq 0$ for two different system sizes  ($N=256$ in pink solid line and $N=512$ in blue dashed line). For -$1<k<0$, from our numerics we find that $a_k - b_k/2$ is slightly different from zero, possibly  due to finite size effects, hence we get  a system size dependence in the relative fluctuation as well. (Inverse temperature $\beta$ is taken to be unity.)
		}
		\label{rel_fluct}
	\end{figure}
	\subsection{Error estimation in the exponent $b_k$}
	Note that $b_k$ is extracted from variance data for last two largest system sizes. Here we perform error analysis of the exponent value $b_k$ for different $k$ in Fig.~3 of the main text. To do that we first express the exponent $b_k$ in terms of the variance $\mathcal{V}(N)=\sigma^2_{\Delta_{N/2}}$ for a given system size $N$:
	\beq
	b_k=\frac{\ln(\mathcal{V}({N/2}))-\ln(\mathcal{V}({N}))}{\ln(2)}.
	\eeq
	The maximum fractional error in the computation of $b_k$ can be written  in terms of the 
	error $\delta \mathcal{V}({N/2})$ in the variance $\mathcal{V}({N})$ as 
	\begin{align}
	(\delta b_k)_{max} & = \left ( \frac{\delta\mathcal{V}({N/2})}{\mathcal{V}({N/2})}+\frac{\delta \mathcal{V}({N})}{\mathcal{V}({N})} \right ) \times \frac{1}{\ln(2)}.
	\label{error_bk}
	\end{align}
	The error  in the variance  is given by 
	\begin{align}
	\delta \mathcal{V}({N}) \approx  \frac{\sqrt{\langle (\Delta_{N/2}-\langle \Delta_{N/2}\rangle)^4\rangle - \langle (\Delta_{N/2}- \langle \Delta_{N/2}\rangle)^2 \rangle^2}}{\sqrt{R}},
	\end{align}
\begin{widetext}
	\begin{figure}[t]
	\begin{minipage}[H]{1.0\textwidth}
		\centering		\includegraphics[width=\textwidth]{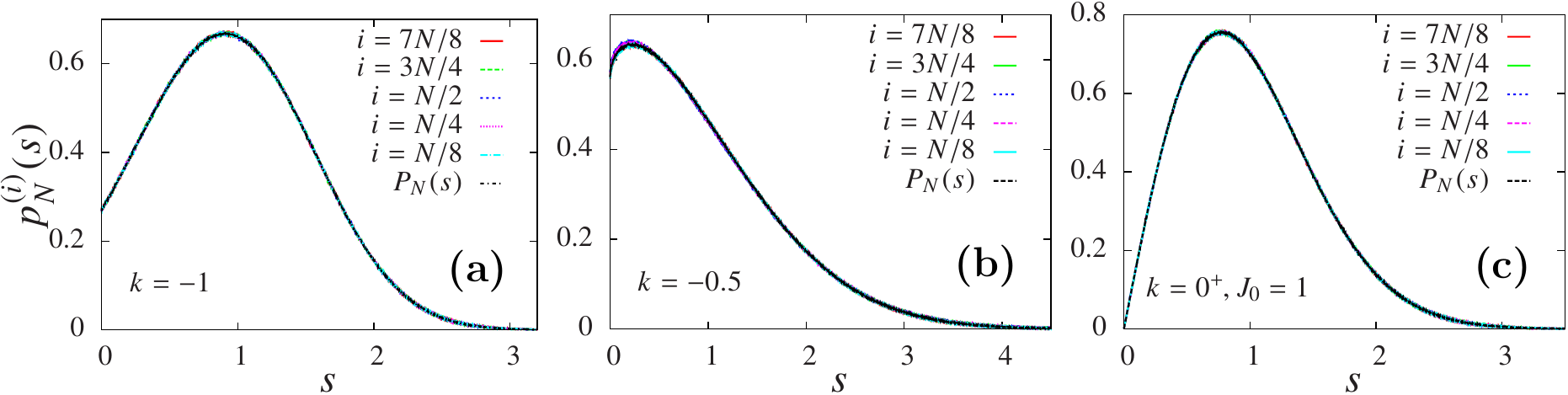}		
		\caption{The distributions of the $i$-th normalized gap $p^{(i)}_N(s)= \langle  \delta (s - s_i) \rangle$ are shown in the regime $-1 \leq k \leq 0$. It is clear that  $p^{\rm (i)}_N(s)$ is independent of $i$. In these figures we have taken $N=256$ at inverse temperature $\beta=1$. 
						}
		\label{Fraction}
	\end{minipage}
\end{figure}
\end{widetext}
	where $R$ is the number of independent realisations. We choose $R$ independent realisations in the simulation by leaving out some Monte-Carlo cycles between successive data collections so as to ensure negligible correlation. We then use the above expressions to compute the error bars in Figs. 2 and 3 in the main text.
	\section{Interesting aspects of the gap statistics for $-1\leq k \leq 0$}
	\label{Interesting}
	The system size dependence of relative fluctuations $\frac{\sigma_{\Delta_i}}{\langle \Delta_i \rangle}$ in the bulk is characterized by the exponent $(a_k-b_k/2)$. From the conjecture Eq.~\eqref{eq:bk} the exponent $(a_k-b_k/2)$ is zero throughout the range $-1 \leq k \leq 0$. We believe that the slight deviations from the predictions for few values of $k$ are due
	to finite size effects. Interestingly, in the range $-1 \leq k \leq 0$, we find that the relative fluctuations have a weak dependence on $i$, in the bulk (with almost independent for $k=-1,0^+$.  This is shown in Fig:~\ref{rel_fluct}. 
	We find that, not only do the relative fluctuations $\frac{\sigma_{\Delta_i}}{\langle \Delta_i \rangle}$ have small variations with $i$, but also the spacing distributions $p^{\rm (i)}_N(s)= \langle  \delta (s - s_i) \rangle$ for individual gaps are also almost the same (see Fig.~\ref{Fraction}). However we expect the differences between $P_N(s)$ and $P^{i}_N(s)$, for $i$ in bulk, to show  up at very small and very large $s$ where edge statistics (e.g  $i =1$) could start dominating over bulk behaviour.
		\begin{figure}[h]
			\begin{minipage}[H]{0.49\textwidth}
				\centering		\includegraphics[width=\textwidth]{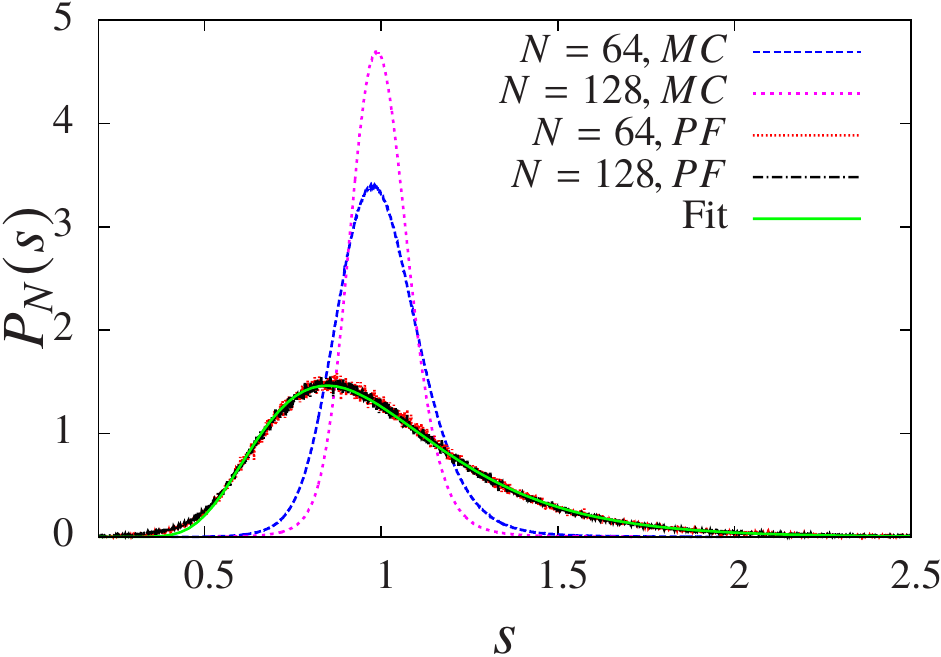}		
			\end{minipage}
			\caption{ Comparison of the level spacing distribution using `picket fence (PF)' approximation and the actual distribution using Monte-Carlo simulation (MC). The distribution using PF approximation is fitted with the Eq.~\eqref{Dstrbn_BG} (Fit) with parameters 
				$A=410.354,B=1.160,C=4.441$. We observe that there is no convergence of the actual level spacing distribution (MC) with system size. ( Inverse temperature $\beta$ is taken to be unity.) 
			}
			\label{comparison_GWS}
		\end{figure}
	\section{Calogero-Moser system}
	\label{sec:CM_supp}
	For the CM model (k = 2) it
	was claimed~\cite{bogomolny_PRL_2009} that the spacing distribution follows a
	form analogous to the WS, namely,
	\beq
	P(s)=Ae^{-B^2/s^2-C s},
	\label{Dstrbn_BG}
	\eeq
	where $B=1.46$ is {$a$} fitting parameter {while} $A$ and $C$ are fixed from the two equations:
	$$
	\int_{0}^{\infty} P(s)ds=1 \text{~and~} \int_{0}^{\infty} sP(s)ds=1.
	\label{Constraint_BG}$$
	However, our results for this system  differ from this claim. For the CM model, we found that the distribution,  $P_N(s)$, of $s$  does not in fact converge with increasing $N$, which thus indicates that the generalization of the Wigner surmise does not work here. We believe that the disagreement can be attributed to  the `picket fence' {(PF)} approximation ($x_j \sim j$) used in the numerical computation of the level spacing distribution in  Ref.~\onlinecite{bogomolny_PRL_2009}. Using this approximation, the Lax matrix $\tilde{L}$ takes the form
	\beq
	\tilde{L}_{nm}=p_n \delta_{nm}+ \frac{i}{2} \frac{1- \delta_{nm}}{(n-m)}, 
	\eeq
	where $p_n$ is chosen from a uniform distribution between $-1$ and $1$.  From the eigenvalues of  $\tilde{L}$, the level spacing distribution was obtained and shown to follow the expression Eq.~\ref{Dstrbn_BG}. Using this method we can indeed verify (see Fig.~\ref{comparison_GWS}) that the  level spacing distribution converges to {an} $N$-indepependent form that is described by a WS-like distribution. On comparing  the level spacing distribution obtained using the above approximation  (denoted by PF in Fig.~\ref{comparison_GWS}) with the actual level spacing distributions (denoted by MC), we observe that they are completely different. In particular, for the true distribution, {we find that there is no convergence as the system size grows}.

	\bibliography{Supplementary.bib}
	